\documentclass[aps,prl,twocolumn,floatfix]{revtex4-1}

\usepackage{graphicx}
\usepackage{amsmath,bm}
\usepackage{mathtools}
\usepackage{braket}
\usepackage{bbold}
\usepackage{nicefrac}
\usepackage{multirow}

\newcommand*\Bell{\ensuremath{\boldsymbol\ell}}

\begin{document}

\title{Experimental characterization of singlet scattering channels in long-range Rydberg molecules}

\author{Heiner Sa{\ss}mannshausen}
\author{Fr\'ed\'eric Merkt}
\author{Johannes Deiglmayr}
\affiliation{Laboratory of Physical Chemistry, ETH Zurich, Switzerland}
\email{jdeiglma@ethz.ch}

\newcommand{\AR}{{\textrm{A}}}
\newcommand{\SF}{{$ns\,n'f$}}
\newcommand{\PP}{{$np\,np$}}
\newcommand{\BR}{{\textrm{B}}}
\newcommand{\TS}{{$^3\Sigma$}}
\newcommand{\STS}{{$^{1,3}\Sigma$}}

\begin{abstract}
  We observe the formation of long-range Cs$_2$ Rydberg molecules consisting of a Rydberg and a ground-state atom by photoassociation spectroscopy in an ultracold Cs gas near 6s$_{1/2}$($F$=3,4)$\rightarrow$$n$p$_{3/2}$ resonances ($n$=26-34). The spectra reveal two types of molecular states recently predicted by D. A. Anderson, S. A. Miller, and G. Raithel [Phys. Rev. A \textbf{90}, 062518 (2014)]: states bound purely by triplet $s$-wave scattering with binding energies ranging from 400~MHz at $n$=26 to 80~MHz at $n$=34, and states bound by mixed singlet-triplet $s$-wave scattering with smaller and $F$-dependent binding energies. The experimental observations are accounted for by an effective Hamiltonian including $s$-wave scattering pseudopotentials, the hyperfine interaction of the ground-state atom, and the spin-orbit interaction of the Rydberg atom. The analysis enabled the characterization of the role of singlet scattering in the formation of long-range Rydberg molecules and the determination of an effective singlet $s$-wave scattering length for low-energy electron-Cs collisions.
\end{abstract}

\maketitle

 Atoms in Rydberg states of high principal quantum number $n$ are weakly bound systems and are extremely sensitive to their environment. In 1934, Amaldi and  Segr\`e~\cite{Amaldi1934} observed the pressure-dependent shift and broadening of the Rydberg series of alkali atoms in a gas cell, an effect which was explained by~\citet{Fermi1934} as originating from the elastic scattering between slow Rydberg electrons and ground-state atoms within the Rydberg orbit. He modeled the pressure shift using a pseudopotential
\begin{equation}\label{eq:Vpseudo}
V(R)=2 \pi  a |\Psi(\bm{R})|^2 \;,
\end{equation}
where $a$ is the scattering length and $|\Psi(\bm{R})|^2$ the probability density of the Rydberg electron at the position $\bm{R}$ of the neutral perturber. Measurements of pressure shifts in Rydberg states thus provide information on the cross sections of elastic collisions between slow electrons and atoms and molecules~\cite{Fermi1934,massey1969}. Equation~\eqref{eq:Vpseudo} implies the existence of oscillating interaction potentials between Rydberg and ground-state atoms~\cite{vitrant1982,dePrunele1987}. A manifestation of such potentials are long-range diatomic molecules in which a ground-state atom having a negative $s$-wave scattering length is attached to a Rydberg atom at a distance corresponding to an antinode of $\Psi(\bm{R})$, as was first pointed out by~\citet{Greene2000}. Such molecules were first observed experimentally by \citet{Bendkowsky2009} following excitation of Rb atoms close to $n$s$_{1/2}$ Rydberg states with $n$=35-37. Later investigations led to the detection of long-range Rb$_2$ molecules correlated to $n$p$_{1/2,3/2}$~($n$=7-12)~\cite{Bellos2013}, $n$d$_{3/2,5/2}$~($n$=34-40)~\cite{Anderson2014}, and $n$d$_{3/2,5/2}$~($n$=40-49)~\cite{Krupp2014} dissociation asymptotes and long-range Cs$_2$ molecules correlated to $n$s$_{1/2}$~($n$=31-34)~\cite{Tallant2012} and $n$s$_{1/2}$~($n$=37,39,40)~\cite{Booth2014} asymptotes. The analysis of the experimental data confirmed the overall validity of Eq.~\eqref{eq:Vpseudo}, revealed contributions from triplet $p$-wave scattering channels, and enabled the determination of triplet $s$- and $p$-wave scattering lengths that confirmed theoretical predictions~\cite{Bendkowsky2010}.

Singlet $s$-wave scattering lengths are expected to be either positive or much smaller than triplet scattering lengths~\cite{Fabrikant1986,Bahrim2001} and thus scattering in singlet channels has not been included in the analysis of the experimental data. \citet{Anderson2014a} have recently predicted that the hyperfine coupling in the ground-state atom should effectively mix triplet and singlet scattering channels and lead to shallow bound states in addition to the more strongly bound states resulting from pure triplet scattering.

Here, we report on the observation of these shallow bound states correlated to $n$p$_{3/2}$ Rydberg states in the range $n$=26-34 in an ultracold Cs gas. We show that the well depths of these shallow states depend on the hyperfine component ($F$=3 or $F$=4) in which the ground-state atoms are prepared, model the experimental observations theoretically, and extract an effective singlet $s$-wave scattering length for low-energy $e^-$-Cs collisions from the experimental data.

Following Ref.~\cite{Anderson2014a}, we model the interaction between the Rydberg and the ground-state atom using
\begin{equation}\label{eq:Hmol}
\hat{H}=\hat{H}_0+\hat{H}_\textrm{SO}+\hat{H}_\textrm{HF}+\hat{P}_\textrm{S}\cdot \hat{V}_\textrm{S}+\hat{P}_\textrm{T}\cdot \hat{V}_\textrm{T},
\end{equation}
where $\hat{H}_0$ is the Hamiltonian of the Rydberg atom without spin-orbit interaction, $\hat{H}_\textrm{SO}=A_\textrm{SO}\, \Bell_r\otimes\bm{s}_r$ is the spin-orbit interaction of the Rydberg electron with angular momentum $\Bell_r$ and electron spin $\bm{s}_r$, and $\hat{H}_\textrm{HF}=A_\textrm{HF}\, \bm{i}_g\otimes\bm{s}_g$ is the hyperfine interaction of the ground-state atom with nuclear spin $\bm{i}_g$ and electron spin $\bm{s}_g$. Separate terms for singlet (S) and triplet (T) scattering are obtained by using the projection operators $\hat{P}_\textrm{S}=-\bm{s}_r\otimes\bm{s}_g+\frac{1}{4}\cdot\hat{\mathbb{1}}$ and $\hat{P}_\textrm{T}=\bm{s}_r\otimes\bm{s}_g+\frac{3}{4}\cdot\hat{\mathbb{1}}$ and representing the corresponding Fermi-contact-interaction operators by $\hat{V}_i = 2 \pi  a_i \hat{\delta}^3(\bm{r}-R\hat{\bm{z}})$, with $i$=S,\,T, for $s$-wave scattering between the Rydberg electron at position $\bm{r}$ and the ground-state atom located at distance $R$ along the molecular $z$-axis~\cite{Omont1977}.  We neglect the rotational motion of the molecule, because the spacings between the accessible rotational levels are less than their natural linewidths. The hyperfine structure of the Rydberg atom is also neglected, because the corresponding interaction is much weaker than all other relevant interactions~\cite{sasmannshausen2013}. To reproduce experimental binding energies, we adjust the zero-energy limit of the energy-dependent scattering lengths (Ref.~\cite{Fabrikant1986}, Eqs. (1,4-6)). The spin-orbit coupling constants $A_\textrm{SO}$ are extracted from experimental fine-structure splittings~\cite{goy1982} and we use the hyperfine coupling constant $A_\textrm{HFS}$ from Ref.~\cite{Arimondo1977}.

\begin{table}[tb]
\centering
\begin{tabular}{|cc|c|c|c|c|}
\hline
\multicolumn{2}{|c|}{Rydberg asymptote}                              &  Contributions                             & $g_3$ & $g_4$ &  Binding via\\  \hline
\multirow{2}{*}{$n$p$_{1/2}$} & \multirow{2}{*}{$|\omega_r| = 1/2$}  &  $^1\Sigma^+$, $^3\Sigma^+$, $^1\Pi$, $^3\Pi$ & 8 &  8  & $a_\mathrm{S}$, $a_\mathrm{T}$ \\  \cline{3-6}
                              &                                      &  $^3\Sigma^+$, $^1\Pi$, $^3\Pi$               & 6 &  10 & $a_\mathrm{T}$\\  \hline
\multirow{3}{*}{$n$p$_{3/2}$} & \multirow{2}{*}{$|\omega_r| = 1/2$}  &  $^1\Sigma^+$, $^3\Sigma^+$, $^1\Pi$, $^3\Pi$ & 8 &  8  & $a_\mathrm{S}$, $a_\mathrm{T}$ \\  \cline{3-6}
                              &                                      &  $^3\Sigma^+$, $^1\Pi$, $^3\Pi$               & 6 &  10 & $a_\mathrm{T}$\\  \cline{2-6}
                              & $|\omega_r| = 3/2$                   &  $^1\Pi$, $^3\Pi$                             & 14&  18 & --- \\  \hline
\end{tabular}
\caption{\label{tab:couplings}Asymptotes and binding mechanism. First column: fine-structure asymptote of the Rydberg atom. Second column: symmetries contributing to the molecular state. Third and fourth column: degeneracies $g_3$ and $g_4$ of the molecular states associated with Cs\,$n$p$_{j}$,6s$_{1/2}$($F$=3,\,4) asymptotes. Last column: $s$-wave scattering lengths contributing to the binding.}
\end{table}
The Hamiltonian matrix corresponding to Eq.~\eqref{eq:Hmol} is set up in the uncoupled basis $\ket{\ell_r,\lambda_r}\ket{s_r,\sigma_r}\ket{s_g,\sigma_g}\ket{i_g,\omega_{i,g}}$ ($i_g=7/2$ for $^{133}$Cs) for a single $n$p Rydberg state and the energetically closest Rydberg state, ($n$-1)d. The quantum number $\Omega_\textrm{tot}\equiv \lambda_r + \sigma_{r}+\sigma_{g}+\omega_{i,g}$ of the projection of the total angular momentum on the internuclear axis is a good quantum number. The ground-state atom can be in either of the two $F$=3,\,4 ($\bm{F}$$\equiv$$\bm{i}_g$+$\bm{s}_g$) hyperfine levels, and the $n$p Rydberg atom can be in either the $n$p$_{3/2}$ ($\omega_r$$\equiv$$\lambda_r+\sigma_r$=$-3/2$,\,$-1/2$,\,$1/2$,\,$3/2$) or the $n$p$_{1/2}$ ($\omega_r$=$-1/2$,\,$1/2$) spin-orbit levels. The spin-orbit interaction of the Rydberg electron completely mixes the $\lambda_r$=0 and $|\lambda_r|$=1 components of the $|\omega_r|$=$1/2$ levels, whereas the $|\omega_r|$=$3/2$ levels have pure $\Pi$ character. Because we restrict the $e^-$-Cs interaction to $s$-wave scattering, only the $\lambda_r$=0 ($\Sigma^+$) components of the Rydberg wavefunction contribute to the binding~\cite{Omont1977}. The $|\omega_r|$=$3/2$ levels are thus non-bonding and the strength of the bond of the $|\omega_r|$=$1/2$ levels is proportional to their $\lambda_r$=0 ($\Sigma^+$) character. The selection of the $\lambda_r$=0 component in the interaction terms $\hat{V}_\mathrm{S}$ and $\hat{V}_\mathrm{T}$ implies that $\omega_r$ equals $\sigma_r$ in all components of the wavefunctions contributing to binding and that $s$-wave singlet and triplet channels of the $e^-$-Cs scattering interaction are not mixed by $\hat{H}_\textrm{SO}$. These channels can, however, be mixed by $\hat{H}_\textrm{HF}$. Remarkably, the eigenstates of Eq.~\eqref{eq:Hmol} form two distinct subgroups, one containing states which do not possess any $^1\Sigma^+$ character (but $^3\Sigma^+$, $^1\Pi$ and $^3\Pi$) and for which the binding is entirely dictated by the triplet $s$-wave interaction, and one containing states which possess both $^3\Sigma^+$ and $^1\Sigma^+$ (as well as $^1\Pi$ and $^3\Pi$) character and for which the binding results from both singlet and triplet $s$-wave scattering interactions. Table~\ref{tab:couplings} gives an overview of the character and degeneracy factors of the resulting bound states. For simplicity we refer to the states bound via the $^3\Sigma^+$ component as \TS\ states and to those for which the binding results from both $^3\Sigma^+$ and $^1\Sigma^+$ components as \STS\ states. The \TS\ states are those that have been observed in previous experiments~\cite{Bendkowsky2009,Bellos2013,Anderson2014}.

\begin{figure}[b]
\begin{center}
\includegraphics[width=\linewidth]{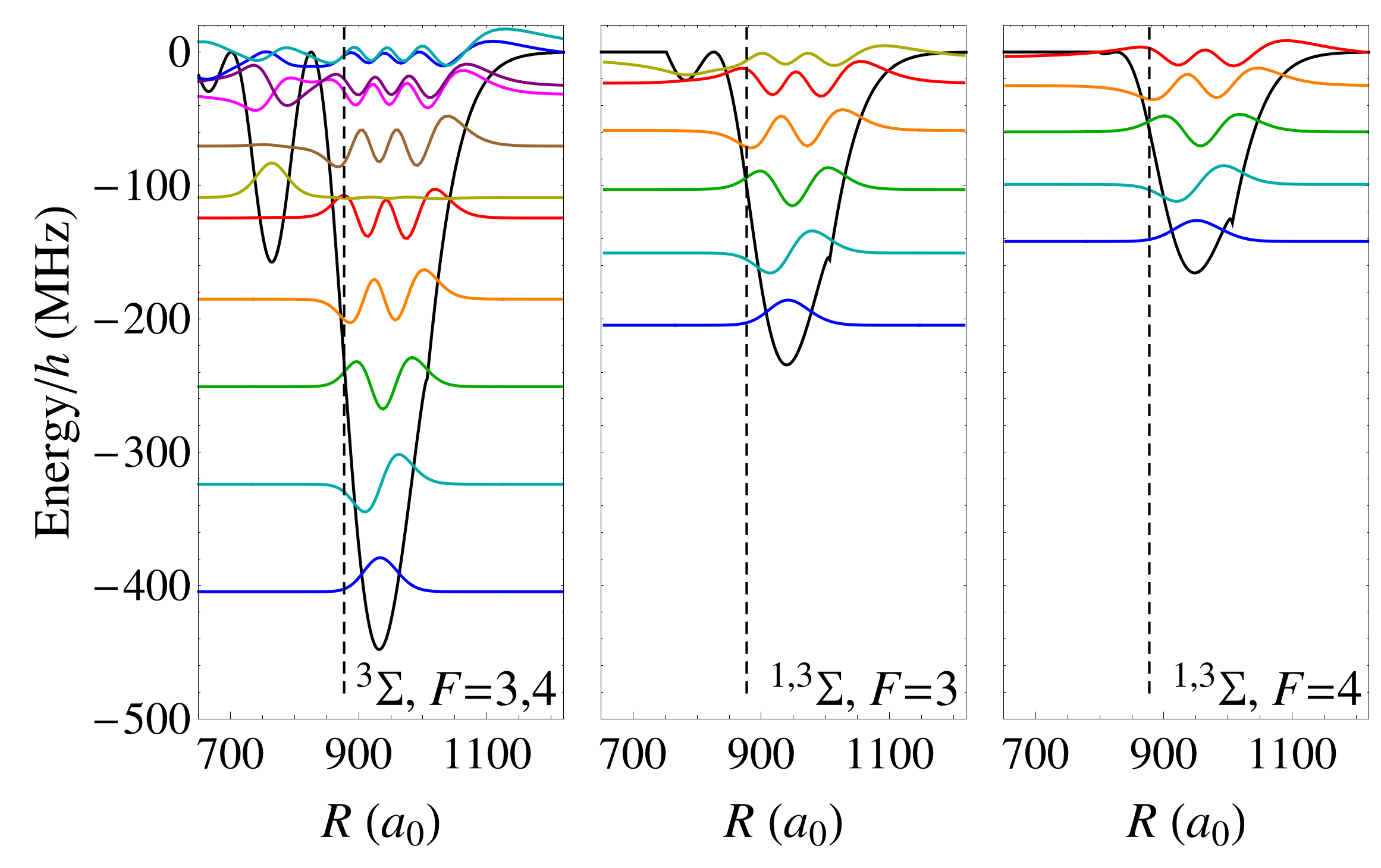}
\caption{\label{fig:calc} (Color online) Potential-energy curves (black lines) and vibrational wave functions (colored lines) for $^3\Sigma$ states dissociating to the $26p_{3/2}$,6s$_{1/2}$($F$=3,4) asymptotes (left panel), $^{1,3}\Sigma$ state dissociating to the $26p_{3/2}$,6s$_{1/2}$($F$=3) (middle panel), and $26p_{3/2}$,6s$_{1/2}$($F$=4) (right panel) asymptote. Vertical dashed lines mark the positions where the semiclassical Rydberg-electron energy equals the energy of the $^3P_0$ Cs-$e^-$ scattering resonance (see text).}
\end{center}
\end{figure}

A Cooper minimum in the 6s$_{1/2}$$\rightarrow$$n$p$_{1/2}$ photoexcitation cross section of Cs~\cite{raimond1978} prevents the observation of states correlated to $n$p$_{1/2}$ asymptotes and we thus focus on the states correlated to $n$p$_{3/2}$ asymptotes. Potential-energy curves (PECs) for the long-range molecules are obtained by determining the eigenvalues of Eq.~\eqref{eq:Hmol} as a function of the separation $R$ between the atoms. We use the mapped Fourier-grid method~\cite{Kokoouline1999} to obtain the vibronic eigenstates of these PECs. The PECs and vibrational wavefunctions of the 26p$_{3/2}$,\TS\ and 26p$_{3/2}$,\STS\ states, calculated for parameters derived from the analysis of our spectroscopic data, are depicted in Fig.~\ref{fig:calc}. The \TS\ PECs do not depend on the hyperfine state of the ground-state atom, while the \STS\ state correlated to the $F$=3 asymptote is more strongly bound than the \STS\ state correlated to the $F$=4 asymptote.

\begin{figure*}[bt]
\begin{center}
\includegraphics[width=0.85\textwidth]{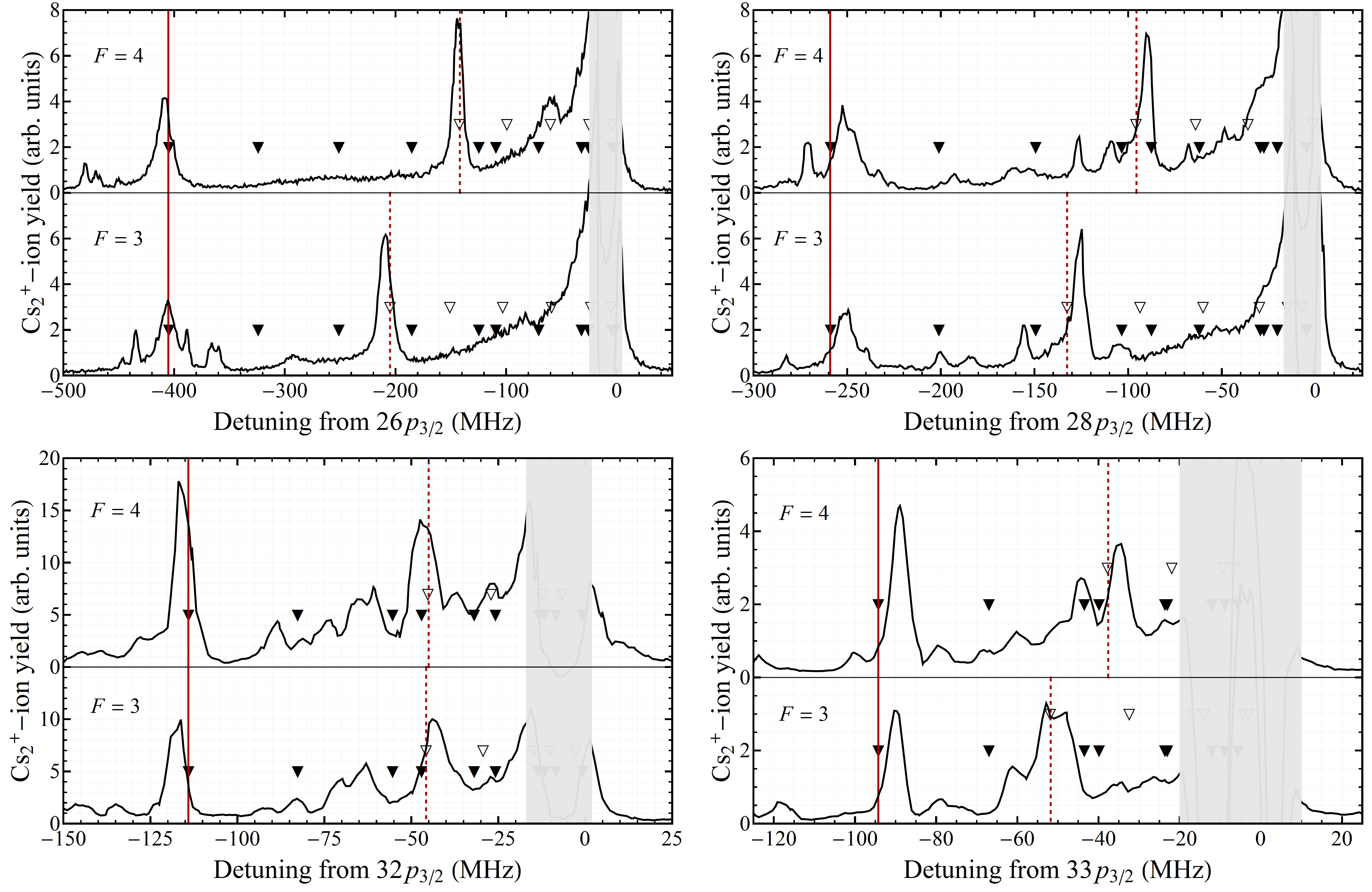}
\caption{\label{fig:allspec}Detected Cs$_2^+$ ions as function of laser frequency near the $n$p$_{3/2}$,6s$_{1/2}$($F$=3,\,4) asymptotes. Solid (dashed) red lines mark the calculated positions of the $v$=0 levels of \TS\ (\STS) states. Calculated positions of higher vibrational levels are also shown (full and open triangles, respectively). The strongly saturated atomic transitions correspond to the grey areas.}
\end{center}
\end{figure*}

The long-range Cs$_2$ molecules are photoassociated from samples of ultracold Cs atoms ($3 \cdot 10^8$ atoms at a density of $2\cdot10^{11}$\,cm$^{-3}$ and a temperature of 40\,$\mu$K), released from a vapor-loaded compressed magneto-optical trap~\cite{Petrich1994}. Stray magnetic and electric fields are compensated to below 2~mG and 20~mV/cm, respectively~\cite{sasmannshausen2013,Deiglmayr2014}. The atoms are optically pumped into one of the two hyperfine states ($F$=3,\,4) of the 6s$_{1/2}$ ground state and excited with a single-mode cw UV laser (100~mW power, 5~MHz linewidth, 40~$\mu$s excitation pulse length) using intensities around 100\,W/cm$^{2}$. The frequency relative to the $n$p$_{3/2}$ atomic resonance is calibrated at an accuracy of 5~MHz using a scanning Fabry-Perot cavity and a frequency-stabilized HeNe laser. After excitation, spontaneously formed ions~\footnote{Comparable spectra are obtained when using field ionization and detection of Cs$^+$ ions.} are detected with a micro-channel-plate detector. On molecular resonances, we observe both atomic Cs$^{+}$ (attributed to Penning ionization) and molecular Cs$_{2}^{+}$ ions (attributed to associative ionization), the relative strength of the former increasing with the density of excited molecules.

\begin{figure}[tb]
\begin{center}
\includegraphics[width=\linewidth]{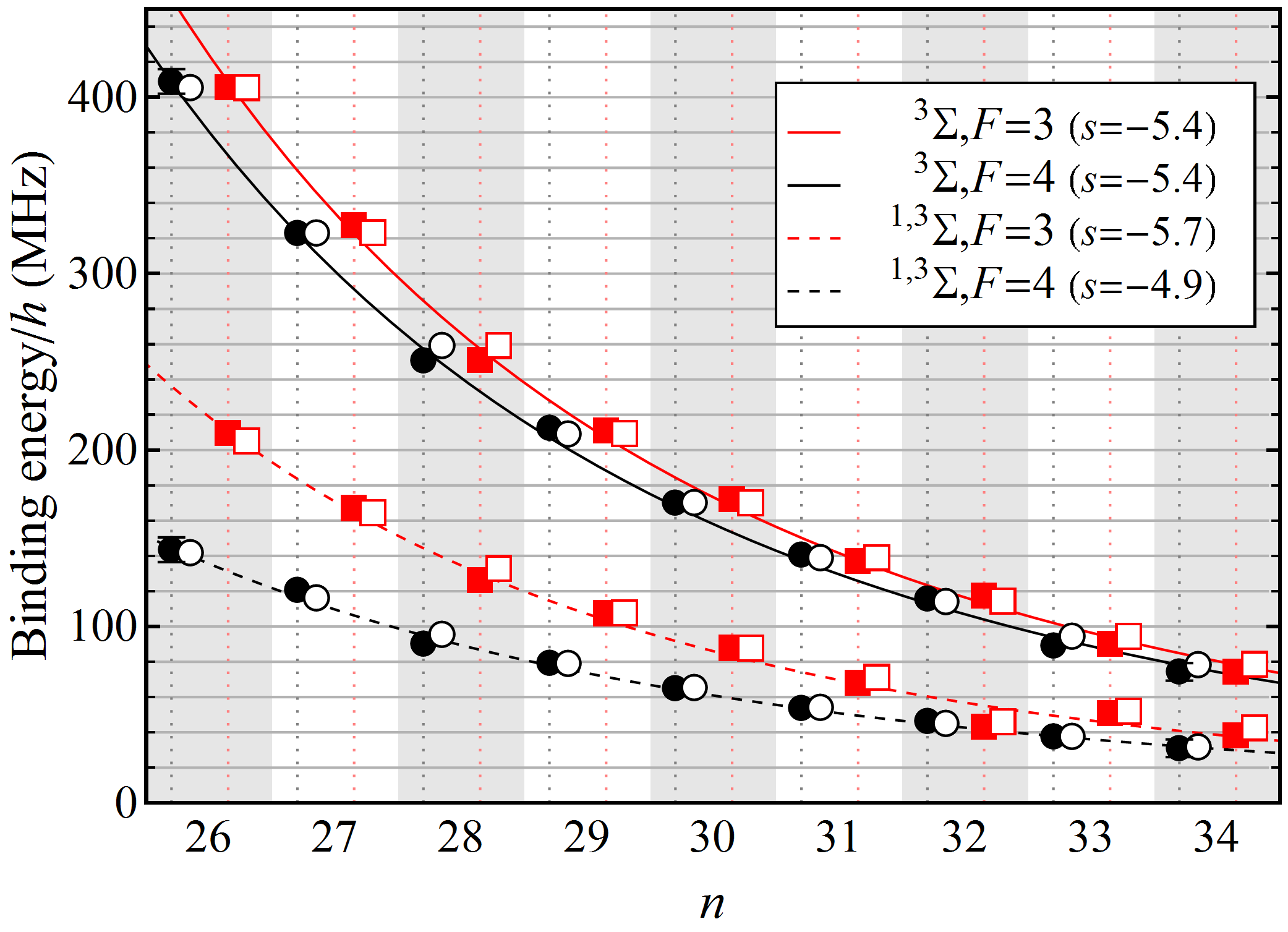}
\caption{\label{fig:allpos}
Comparison of experimental (full symbols) and calculated (open symbols) binding energies of the \TS($v$=0) and \STS($v$=0) levels. The solid/dashed lines are fits of a ${\left(n^*\right)}^s$ scaling law to the experimental binding energies to guide the eye. Black and red symbols designate states with the ground-state atoms prepared in $F$=4 and $F$=3, respectively.}
\end{center}
\end{figure}

Fig.~\ref{fig:allspec} presents photoassociation spectra of Cs atoms prepared in the $F$=3 and $F$=4 hyperfine levels to long-range Cs$_2$ molecules correlated to $n$p$_{3/2}$ Rydberg states ($n$=26, 28, 32, and 33). Several photoassociation resonances are observed on the low-frequency side of the strongly saturated atomic 6s$_{1/2}$($F$=3,4)$\rightarrow$$n$p$_{3/2}$ resonances. The two most intense photoassociation resonances are assigned to the vibronic ground state of the \TS\ and \STS\ states. We determine the binding energies of the \TS($v$=0) and \STS($v$=0) levels for all $n$ values between 26 and 34. In Fig.~\ref{fig:allpos} these energies are compared to binding energies calculated with the model outlined above after adjustment of the zero-energy $s$-wave singlet ($a_{\textrm{S},0}$) and triplet ($a_{\textrm{T},0}$) scattering lengths, to $a_{\textrm{T},0}=-21.8\pm0.2\;a_0$ and $a_{\textrm{S},0}=-3.5\pm{0.4}\;a_0$, in a least-squares fit~\footnote{If we model the energy dependence of the scattering lengths by $a(k)=a_\textrm{0}+\frac{\pi}{3}\alpha k(R)$ with the ground-state atom polarizability $\alpha=402.2$~a.u. and the semiclassical electron momentum $k=\sqrt{2/R-1/{n^*}^2}$~\cite{Omont1977}, we find $a_{\textrm{T},0}=-20.5\pm0.2\;a_0$ and $a_{\textrm{S},0}=-3.7\pm{0.4}\;a_0$.}. The fact that the adjustment of only two parameters permits the reproduction of 36 resonances within their experimental uncertainty of 5~MHz leaves no doubt concerning the validity of the assignments. As predicted by the model calculations (see also Fig.~\ref{fig:calc}), the positions of the \TS($v$=0) levels are identical for $F$=3 and $F$=4 within the experimental uncertainty whereas ($F$=4),\STS($v$=0) levels are less strongly bound than the ($F$=3),\STS($v$=0) levels. An exception is found at $n$=32, where the \STS($v$=0) levels correlated to 32p$_{3/2}$,6s$_{1/2}$($F$=3) and 32p$_{3/2}$,6s$_{1/2}$($F$=4) have almost identical binding energies. This irregularity is well described by our model, which indicates that it originates from an accidental perturbation of the 32p$_{3/2}$,6s$_{1/2}$($F$=3) state by the almost degenerate 32p$_{1/2}$,6s$_{1/2}$($F$=4) state. Indeed, at $n=32$, the $n$p$_{1/2}$-$n$p$_{3/2}$ fine-structure splitting, which scales as ${n^*}^{-3}$, almost exactly matches the ground-state hyperfine splitting. The dependence of the \TS($v$=0) binding energies on the effective quantum number $n^*$ deviates from the previously encountered ${n^*}^{-6}$ scaling~\cite{Bendkowsky2009} (see Fig.~\ref{fig:allpos}). Our calculation shows that this deviation is caused by the increasing contribution of higher-order, energy-dependent terms of the scattering length for lower $n$.

Our values of the triplet and singlet $s$-wave scattering lengths are effective, model-based values obtained assuming that $p$-wave scattering is negligible. This assumption is fulfilled for low electron energies and in the absence of scattering resonances. The lowest $p$-wave scattering resonance (the $^3P_0$ shape resonance in Cs$^-$~\cite{Scheer1998}) is located at an electron kinetic energy $\frac{p_c^2}{2m_e}$ of 4~meV~\cite{Tallant2012} which is (in a semi-classical description) reached by the Rydberg electron at the critical radius $r_c$ corresponding to $p_c^2=(-\frac{1}{n^{*2}}+\frac{2}{r_c})$, and marked in Fig.~\ref{fig:calc} by vertical dashed lines. Although the minima of the outermost potential wells are located beyond $r_c$ and the $v$=0 levels should not be strongly affected by $p$-wave scattering for $n\geq 26$, Fig.~\ref{fig:calc} indicates that PECs for lower $n$ values and excited vibrational levels of the outermost well could be significantly impacted by $p$-wave scattering channels. We believe this to be the reason for the inability of our effective $s$-wave scattering model to reliably predict the positions of excited vibrational levels, which are marked by triangles in Fig.~\ref{fig:allspec}, and of further resonances observed in our experiments for $n$=19-25.

\citet{Fabrikant1986} and, more recently, \citet{Bahrim2001} calculated the $e^-$-Cs $s$-wave scattering lengths to be $a_{\textrm{T},0}$=$-22.7$~$a_0$ and $a_{\textrm{S},0}$=$-2.4$~$a_0$, and $a_{\textrm{T},0}$=$-21.7$~$a_0$ and $a_{\textrm{S},0}$=$-1.33$~$a_0$, respectively, from an effective-range-theory analysis of higher-energy scattering data. As measure of the deviation between experiment and theory, we consider the absolute error in the zero-energy scattering phase shift $\delta_0$=$\arctan\left(a_i /R_{\mathrm{pol}}\right)+\pi/4$~\cite{Gribakin1993}, where $R_{\mathrm{pol}}$=$\sqrt{\alpha}$=$20.05$~$a_0$ is the typical range of the $e^-$-Cs interaction and $\alpha$ is the polarizability of the Cs ground-state atom. The deviation between experiment and theory is at most 0.1 radian and the agreement can thus be regarded as good. Our fitted model-based values may include effects of higher partial waves in the $e^-$-Cs scattering and interactions with Rydberg states other than the neighboring ($n$-1)d$_{3/2,5/2}$ states in an effective manner. The fact that our model reproduces the experimental observations well over a broad range of $n$ from 26 to 34 suggests that these effects are not dominant.

The interaction operators $\hat{V}_i$ in Eq.~\eqref{eq:Hmol} mix the neighboring $n$p and $(n-1)$d Rydberg states and induce an electric dipole moment in the molecular frame~\cite{li2011,Tallant2012,Anderson2014}. For the 32p$_{3/2}$,6s$_{1/2}$($F$=4),\TS($v$=0) states, our model predicts dipole moments between 21.0 and 25.8~D (mean value 24.4~D) for $|\Omega_{\rm tot}|$ between 1/2 and 9/2. The dipole moments of the 32p$_{3/2}$,6s$_{1/2}$($F$=4),\STS($v$=0) states are smaller and range from -0.5 to 12.0~D (mean value 5.6~D) for $|\Omega_{\rm tot}|$ between 1/2 and 7/2. In an electric field we observe broadenings for these states which are in qualitative agreement with the expected Stark shifts, however the spectra also reveal the strong perturbations due to field-induced state mixing (see appendix) as observed in long-range Rb$_2$ dimers~\cite{li2011}.

\begin{table}[b]
\centering
\begin{tabular}{|c|c|c||c|c||c|c|}
\hline
$n$ & $E_{2,^3\Sigma}^\mathrm{Exp}$ & $E_{2,\dagger}^\mathrm{Exp}$ & $E_{2,^3\Sigma}^\mathrm{Th}$ & $E_{2,^{1,3}\Sigma}^\mathrm{Th}$ &  $E_{2,^{1,3}\Sigma}^*$ & $E_{1,^{1,3}\Sigma}^*$ \\
\hline
35 & -23.0  & -3.8 & -23.1 & -3.9 & -3.6 & -11.0\\
\hline
36 & -19.0  & -2.9 & -19.2 & -3.2 &  -2.9 & -9.1\\
\hline
37 & -16.2  & -2.3 & -16.0 & -2.6 &  -2.4 & -7.5\\
\hline
\end{tabular}
\caption{\label{tab:Rb}Experimental positions $E^\textrm{Exp}_2$ (MHz) of the \TS($v$=0) resonances correlated to $^{87}\textrm{Rb}$,$n$s$_{1/2}$,5s$_{1/2}$($F$=2) asymptotes from Ref.~\cite{Bendkowsky2010}, and calculated positions $E^\textrm{Th}$ of $v$=0 levels in \TS\ and \STS\ states. Also given are $v$=0 binding energies $E_{F,^{1,3}\Sigma}^*$ for $F$=1,\,2 based on optimized values $a_{\textrm{T},0}=-18.7 a_0$ and $a_{\textrm{S},0}=+1.0 a_0$.}
\end{table}
The prominent contributions of the long-range molecules in \STS\ states to the photoassociation spectrum of Cs near the $n$p$_{3/2}$,6s$_{1/2}$ asymptotes implies that these states should also be observable in other systems. In Refs.~\cite{Bendkowsky2009,Bendkowsky2010}, long-range Rb$_2$ molecules were studied by photoassociation of Rb atoms prepared in the 5s$_{1/2}$($F$=2,\,$m_F$=2) level in a magnetic trap. Thus, only the shallow \STS\ states correlating to the $n$s$_{1/2}$,5s$_{1/2}$($F$=2) asymptotes could have been observed. Using a triplet $s$-wave scattering length of $a_{\textrm{T},0}=-18.5 a_0$~\cite{Bendkowsky2009} and the literature value for $a_{\textrm{S},0}=+0.627 a_0$~\cite{Bahrim2001} we predict the positions of the $v$=0 levels of the $n=35-37$ \STS\ shallow potential wells listed in the column $E_{2,^{1,3}\Sigma}^\mathrm{Th}$ of Tab.~\ref{tab:Rb}. Near these positions, all spectra presented in Fig.~2 of Ref.~\cite{Bendkowsky2010} show features marked by dashed lines at positions given in the column $E_{2,\dagger}^\mathrm{Exp}$ of Tab.~\ref{tab:Rb}. After a small adjustment of $a_{\textrm{S},0}$ to 1.0~$a_0$, our calculations of the positions of the \STS($v$=0) levels for $F$=2 match all experimental peak positions within 0.2~MHz (see column $E_{2,^{1,3}\Sigma}^*$ in Tab.~\ref{tab:Rb}). In Ref.~\cite{Bendkowsky2010}, these resonances were, however, attributed to a purely atomic transition,
the shift being the electron-spin Zeeman shift of the $n$s$_{1/2}$ Rydberg level. Our calculations for the \STS($v$=0) levels associated with the $F$=1 component of the ground-state level predict larger binding energies (see column $E_{1,^{1,3}\Sigma}^*$ in Tab.~\ref{tab:Rb}) and we hope that our predictions will be confirmed by future experiments with ultracold Rb samples prepared in the $F$=1 hyperfine component.

In previous studies of \TS\ long-range Rb molecules correlated to $n$s$_{1/2}$ Rydberg states, not only dimers, but also trimers and even larger Rb$_m$ molecules have been observed~\cite{Gaj2014}, the binding energies being proportional to the number ($m-1$) of ground-state atoms. We have not observed any long-range molecules consisting of more than two atoms. We attribute this non-observation to a reduction of the excitation probability of $m$=3 long-range Rydberg molecules by a factor of $(4\pi)^{-1}$ resulting from the change from spherical to linear geometry (imposed by the anisotropy of the $p$-orbitals) when going from $n$s to $n$p Rydberg levels.

\begin{acknowledgments}
 This work is supported financially by the Swiss National Science Foundation under Project Nr.~200020-146759, and the NCCR QSIT, and the EU Initial Training Network COHERENCE under grant FP7-PEOPLE-2010-ITN-265031.
\end{acknowledgments}

%

\newpage

\onecolumngrid
\appendix

\setcounter{figure}{0}
\renewcommand{\thefigure}{S\arabic{figure}}

\section{A. Spectra of molecular states in varying electric fields}

\begin{figure}[bh!]
\begin{center}
\includegraphics[width=0.5 \linewidth]{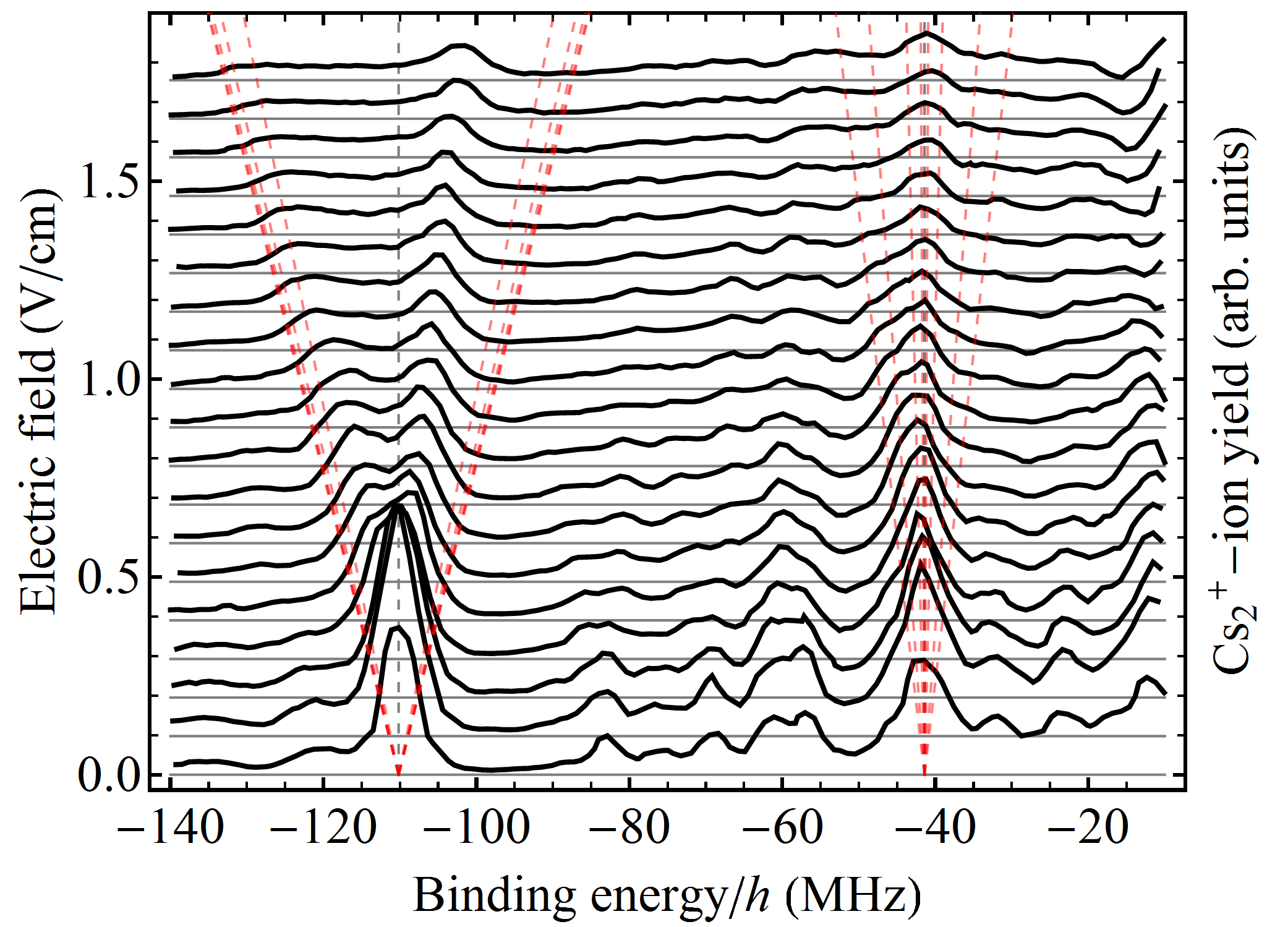}
\caption{Cs photoassociation spectra in different electric fields near the 32p$_{3/2}$,6s$_{1/2}$($F$=4) asymptote. Dashed red lines mark the maximal linear Stark shift of the \TS($v$=0) and \STS($v$=0) levels from calculated molecular electric dipole moments (for all values of $|\Omega_{\rm tot}|$). The quadratic Stark shift of the atomic 32p$_{3/2}$ asymptote (\textit{e.g.}, 50~MHz at 1.5~V/cm) was subtracted from the experimental frequencies.}\label{fig:stark}
\end{center}
\end{figure}

In an externally applied electric field, the widths of the molecular resonances identified as \TS($v$=0) and \STS($v$=0) increase linearly with the magnitude of the electric field $E$, as illustrated in Fig.~\ref{fig:stark} for the 32p$_{3/2}$,6s$_{1/2}$($F$=4) resonances. The molecular dipole moments of the \TS\ and \STS\ states arise in our model from the mixing with the neighboring ($n$-1)d$_{3/2,5/2}$ states induced by the Fermi-contact interaction terms in the Hamiltonian. For the 32p$_{3/2}$,6s$_{1/2}$($F$=4),\TS($v=0$) states, our model predicts dipole moments ranging from 21.0 to 25.8~D (mean value 24.4~D) for $|\Omega_{\rm tot}|$ values between 1/2 and 9/2. The dipole moments predicted for the 32p$_{3/2}$,6s$_{1/2}$($F$=4),\STS($v=0$) states are smaller and range from -0.5 to 12.0~D (mean value 5.6~D) for $|\Omega_{\rm tot}|$ values between 1/2 and 7/2. The polarization of the excitation laser (linear and parallel to the DC electric field) favors the excitation of molecules which are aligned parallel to the electric field axis and one expects the zero-field resonance to evolve in a double-peak structure with a splitting of $\Delta\nu\simeq2 d_{\rm mol} E$ (dashed lines in Fig.~\ref{fig:stark}), where $d_{\rm mol}$ is the molecular dipole moment. While the observed broadenings are in qualitative agreement with the predicted Stark shifts, the asymmetric line shape of the 32p$_{3/2}$,6s$_{1/2}$($F$=4),\TS($v$=0) level is not captured by our model and would require a diagonalization of the interaction Hamiltonian including the interaction with the external electric field and consideration of the rotational structure~\cite{li2011}.

\end{document}